\documentclass[12pt]{article}
\usepackage[pctex32]{graphics}
\begin{document}
\vspace{1cm}
\begin{center}
~
\\
~
{\bf  \Large Statistical Interparticle Potential between Two Anyons }
\vspace{1cm}

                      Wung-Hong Huang\\
                       Department of Physics\\
                       National Cheng Kung University\\
                       Tainan, Taiwan\\

\end{center}
\vspace{2cm}
The density matrix of a two-anyon system is evaluated and used to investigate the ``statistical interparticle potential" following the theory of Uhlenbeck. The main purpose is to see how the statistical potential will depend on the fractional statistical parameter $\alpha$. The result shows that the statistical potential for a two-anyon system with $\alpha\ge {1\over2}$ is always repulsive. For the system with $0<\alpha<  {1\over2}$, the potential is repulsive at short distances and becomes attractive at long distances. It remains only in the boson system ($\alpha=0$) that the repulsive potential arising from the exclusion principle can disappear and lead to an attractive potential at all distances.
\vspace{2cm}
\begin{flushleft}
Physical Review B52 (1995) 15090
\\
*E-mail:  whhwung@mail.ncku.edu.tw\\
\end{flushleft}
\newpage
In recent years, "anyons" which are considered to obey ``fractional" statistics [1,2] have been a subject of intense study. Anyons have found applications in the fields of the fractional quantized Hall effect [3] and high-temperature superconductivity [4]. Most studies have been done in the context of many-body quantum mechanics. In a recent paper [5], Wu derived the occupation-number distribution function of the anyon gas to formulate the theory of quantum statistical mechanics [6,7].  He found the equation of state for the any on
gas in the Boltzmann limit,
$$PV = NkT[1+(2\alpha-1)n\lambda^2/4V],. \eqno{(1)}$$
where $\lambda = \sqrt{2\pi \hbar^2/mkT}$ is the thermal wavelength. In Eq.
(1), the effective "statistical interactions" are attractive if $\alpha< {1\over2}$, but repulsive if $\alpha> {1\over2}$. As this may lead to an impression that the ``statistical interparticle potential" introduced by Uhlenbeck [8] is attractive for the anyon with $\alpha< {1\over2}$ and repulsive for $\alpha> {1\over2}$, we present this paper to show that this is not correct.

Our result shows that the statistical potential for the two-anyon system with $\alpha\ge {1\over2}$ is always repulsive which agrees with Wu's result. But, for the system with $0<\alpha<  {1\over2}$, the potential which appears repulsive at short distance will become attractive at long distances. Only in the boson system ($\alpha = {1\over2}$) can the repulsive potential arising from the exclusion principle totally vanish and thus the statistical potential becomes attractive at all distances. Notice that this does not mean that our result is in contradiction to that in Eq. (1). The reason is that Eq. (1) is the result from the statistical average of the
N-anyon system, while our analysis is only for the two-anyon system and has not yet taken the statistical ensemble average into account.

Following Uhlenbeck's prescription, we first calculate the diagonal elements of the Boltzmann factor matrix for the system of two-anyon gases. Using the wave function of the two-anyon system [1,9] we have
$$<\vec R,r,\theta|e^{-\beta H}|\vec R,r,\theta> = \sum_{ell}\int d\vec K dk <\vec R,r,\theta| exp[-\beta(K^2+k^2)\hbar^2/m]|\vec R,r,\theta>\hspace{3cm}$$
$$={1\over \pi^2 }\int d\vec K exp(-\beta K^2\hbar^2/m) \sum_{\ell}\int dk exp(-\beta k^2)\hbar^2/m)~J^2_{|\ell+\alpha|}(kr)\eqno{(2)}$$
$$= {m^2\over 4\pi^2\hbar^4\beta^2}\sum_\ell exp(-mr^2/2\beta\hbar^2)I_{|\ell+\alpha|}(mr^2/2\beta\hbar^2),\hspace{2cm}\eqno{(3)}$$
where $\vec R$ and $\vec K$ denote the center-of-mass coordinate and momentum, $(r,0)$ the relative coordinate, and $(k,\ell)$ the relative linear momentum and angular momentum, respectively. Without losing generality we will take the angular momentum $\ell$ to be even hereafter, and thus $\alpha =0$ corresponds to the boson gas and $\alpha =1$ the fermion one [1,9]. All the formulas used to perform the evaluations in this paper can be found in Ref.
10.

Under this convention, we can write Eq. (3) in the following form:
$$<\vec R,r,\theta|e^{-\beta H}|\vec R,r,\theta> = {1\over V^2}\sum_\ell exp(-\pi r^2/\lambda^2)I_{|\ell+\alpha|}(\pi r^2/\lambda^2),\eqno{(4)}$$
where $V$ denotes the two-dimensional volume. Then, following the theory of Uhlenbeck [8],  we define the summation term in the above equation as $exp(-\beta v_s)$, where $v_s$, is the so called "statistical interaction potential," i.e.,
$$\beta v_s = - ln\left[ 2\sum_{n=-\infty}^{\infty} exp(-\pi r^2/\lambda^2) I_{|\ell+\alpha|}(\pi r^2/\lambda^2)\right].\eqno{(5)}$$

For the cases of boson ($\alpha =0$) and fennion ($\alpha =1$) the above
equation can be calculated exactly and we recover Uhlenbeck's result, $\beta v_s = - ln[1\pm exp(-2\pi r^2/\lambda^2)]$. This potential reveals an interesting statistical behavior: the negative statistical potential in the boson gas will lead them to bunch together; on the other hand, the positive statistical potential
in the fermion gas tends to push them from each other.

For the two-anyon system which obeys fractional statistics, i.e., $\alpha$  is not an integer, Eq. (5) cannot be reduced to any simple form and we present in Fig. 1 the numerical result of  ``statistical interaction potential" for  $\alpha$ =$0$, $0.1$, $0.2$,$0.5$, and $1$. To confirm our results we will analyze the asymptotic behaviors of the summation term in Eq. (5).
\\
\\
\scalebox{1}{\hspace{5cm}\includegraphics{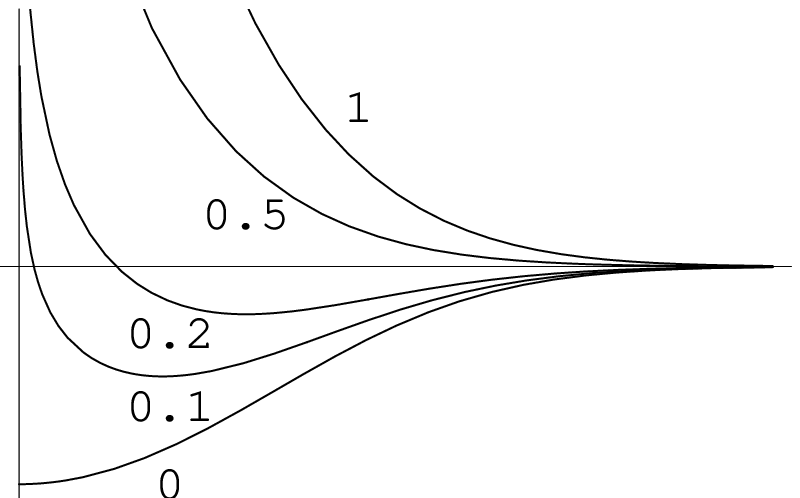}}
\\
{\hspace{3cm} {\it Figure 1.  Statistical interparticle potential $\beta v_s (r/\lambda$) between a pair of anyons with statistical parameter $\alpha$. The value around a curve denotes the value of $\alpha$ associated with it.}
\\
\\
Let us first consider the case of small relative distance. For small $r$, we can approximate the summation term as
$$\sum_{n=-\infty}^{\infty} I_{|2n+\alpha|}(z) e^{-z} \approx \sum_{n=-\infty}^{\infty}[\Gamma(|2n+\alpha|)]^{-1} (2z)^{|2n+\alpha|} e^{-z}$$
$$ \approx [\Gamma(|\alpha+1|)]^{-1} (2z)^{\alpha}+ [\Gamma(3- \alpha)]^{-1} (2z)^{2-\alpha}e^{-z},$$
 where $z = \pi r^2/\lambda^2$. This leads to the following statistical potential at small $r$ [10].
$$\beta  v_s \approx \left\{\begin{array}{c} - ln[2(\pi r^2/2\lambda^2)^\alpha/\Gamma(\alpha+1)].~~~as~r <<\lambda,~~\alpha >0\\
- ln(2-2\pi r^2/2\lambda^2).~~~as~ r <<\lambda,~~\alpha =0.\end{array}\right.\eqno{(6)}$$
Equation (6) shows that all the fractional statistical anyon as well as the fermion gas have the repulsive potential at the short distance. Only the boson gas has the attractive potential at the short distance.

Next, we consider the case of large relative distance $r$. Using the integral representation of the modified Bessel function and Euler-MacLaurin integration formula [10], we obtain the asymptotic form:
$$\sum_{n=-\infty}^{\infty} I_{|2n+\alpha|}(z)=\sum_{n=0}^{\infty} I_{2n+\alpha}(z)+\sum_{n=1}^{\infty} I_{2n-\alpha}(z)\hspace{8cm}$$
$$= {1\over \pi} \int_0^\pi d\phi e^{z cos\phi}\left[\sum_{n=0}^{\infty} cos(2n+\alpha)\phi + \sum_{n=1}^{\infty} cos(2n-\alpha)\phi\right]\hspace{4cm}$$
$$ - {1\over \pi} \int_0^\infty dt e^{-z coth t}\left[\sum_{n=0}^{\infty} e^{-(2n+\alpha)t}sin(2n+\alpha)\phi + \sum_{n=1}^{\infty}  e^{-(2n+\alpha)t} sin(2n-\alpha)\phi\right]$$
$$\approx  \int_0^\pi d\phi e^{z cos\phi}(e^{i\alpha\phi} +e^{i\alpha\phi})[\delta(2\phi)+\delta(2\phi-2\pi)] - {1\over \pi} \int_0^\infty dt e^{-z coth t}\hspace{2cm}$$
$$\hspace{2cm}\left[{1\over2}(e^{-\alpha t} -e^{\alpha t})sin\alpha\pi + {1\over 2(t^2+\pi^2)}[e^{-\alpha t}(\pi cos\alpha \pi + t sin \alpha \pi) + e^{\alpha t}(\pi cos\alpha \pi - t sin \alpha \pi)]\right]$$
$$\approx {1\over2} e^ t (1+e^{-2t}cos\alpha \pi) -{e^{-t}\over \pi} \int_0^\infty dt e^{-zt^2/2}\left[{1\over\pi}cos\alpha \pi - \alpha t sin\alpha \pi\right]\hspace{2.8cm}$$
$$\approx {1\over2} e^ t (1+e^{-2t}cos\alpha \pi) + {e^{-t}\over \pi\sqrt{2z}}\left[\alpha  sin\alpha z- {1\over\pi}cos\alpha \pi\right]\hspace{4.6cm}$$
Thus the statistical potential can be found as
$$\beta  v_s \approx \left\{\begin{array}{c}(cos\alpha \pi)exp(-2\pi r^2/\lambda^2).~~~as~r >>\lambda,~~\alpha \ne {1\over2}\\

[(2 \pi)^{3/2}(r/\lambda)]^{-1}exp(-2\pi r^2/\lambda^2).~~~as~ r >>\lambda,~~\alpha ={1\over2}.\end{array}\right.\eqno{(7)}$$

This result shows that the anyons with statistical parameter $\alpha <{1\over2}$ can exhibit a tendency to bunch together, if the relative distance between two anyons is large. On the other hand, the anyons with statistical parameter $\alpha \ge {1\over2}$ will exhibit a tendency of exclusiving each other at all distances. In all cases, the statistical interparticle potential vanishes exponentially as
$r$ becomes larger than the thermal wavelength $\lambda$, and the semion ($\alpha ={1\over2}$ ) has the shortest range of statistical potential as
can be seen from Eq. (7).

In conclusion, we find that the anyons with statistical parameter $\alpha \ge {1\over2}$ will have positive statistical interparticle potential for all distances. However, for the anyons with $0<\alpha < {1\over2}$, the statistical interparticle potential is repulsive at short distance and becomes attractive at large distances. It remains only in the boson system ($\alpha = {1\over2}$) that the repulsive potential arising from the exclusion principle vanishes and the statis-
tical potential is attractive at all distances.
\\
~
\\
~
\\
~
\\
~
{\bf  \Large References}
\begin{enumerate}
\item  J. M. Leinaas and J. Myrheim, Nuovo Cimento B 38, 1 (1977).
\item  Wilczek, Phys. Rev. Lett. 48, 1144 (1982); 49, 957 (1982).
\item R. B. Laughlin, Phys. Rev. Lett. 50, 1359 (1983); Phys. Rev. B 27, 3383 (1983); B. I. Halperin, Phys. Rev. Lett. 52. 1583 (1984); 52, 2390(E) (1984); R. Prange and S. M. Girvin, The Quantum Halt Effect (Springer, Berlin, 1987).
\item P. B. Wiegmann. Phys. Rev. Lett. 60, 821 (1988); R. B. Laughlin, ibid. 60, 2677 (1988); Y.-H. Chen, F. Wilczek, E. Witten. and B. I. Halperin, Int. J. Mod Phys. B 3, 1001 (1989).
\item 'Y.-S. Wu, Phys. Rev. Lett. 73. 922 (1994); F. D. M. Haldane, ibid.
67, 937 (1991); M. D. Johnson and G. S. Canright. Phys. Rev. B 49, 2947 (1994).
\item C. Nayak and F. Wilczek, Phys. Rev. Lett. 73, 2740 (1994); M. V. N. Murthy and R. Shankar, ibid. 73, 3331 (1994); A. K. Rajagopal, ibid. 74, 1048 (1995).
\item W. H. Huang, ``Boson-Fermion Transmutation and Statistics of Anyon," Phys. Rev. E51 (1995) 3729  [hep-th/0308095].
\item G. E. Uhlenbeck and L. Cropper, Phys. Rev. 41, 79 (1932); R. K. Pathria, Statistical Mechanics (Pergamon, London, 1972).
\item R. Mackenzie and F. Wilczek, Int. J. Mod Phys. A 3, 2827 (1988).
\item  L S. Gradshteyn and I. M. Ryzhik, Table of Integrals, Series, and Products (Academic, New York, 1980).
\end{enumerate}
\end{document}